\documentclass[useAMS,usenatbib,fleqn]{mn2e}  
\usepackage{times}
\usepackage{graphicx}
\usepackage{amsmath}
\usepackage{amssymb}
\usepackage{psfig}
\usepackage{multirow}
\usepackage{cases}

\def\Msun{\hbox{$\rm\thinspace M_{\odot}$}}

\def\xmm{{\it XMM-Newton}}

\def\Mbh{\hbox{$M_{\rmn{BH}}$}}
\def\xmm{{\it XMM-Newton}}

\def\gsim{\mathrel{\hbox{\rlap{\hbox{\lower4pt\hbox{$\sim$}}}\hbox{$>$}}}}
\def\lsim{\mathrel{\hbox{\rlap{\hbox{\lower4pt\hbox{$\sim$}}}\hbox{$<$}}}}

\newcommand{\sinc}{\mathrm{sinc}}
   \title[The flux-dependent time lags in NGC 4051]{The flux-dependent X-ray time lags in NGC 4051}
   \author[W. N. Alston et al.]{W. N. Alston$^{1}$\thanks{wna3@leicester.ac.uk},
        S. Vaughan$^{1}$, P. Uttley$^{2}$\\
        $^{1}$University of Leicester, X-ray \& Observational Astronomy Group, Department of Physics and Astronomy, Leicester, LE1 7RH. \\
        $^{2}$Astronomical Institute ``Anton Pannekoek'', University of Amsterdam, Science Park 904, 1098XH, Amsterdam, the Netherlands}

\date{Accepted 2013 July 24. Recieved 2013 July 22; in original form 2013 February 24}
 
\pagerange{\pageref{firstpage}--\pageref{lastpage}} \pubyear{2013}

\begin{document}
\label{firstpage}
\maketitle

\begin{abstract}
We present an analysis of the X-ray time lags for the highly variable Seyfert 1 galaxy NGC 4051, based on a series of \xmm\ observations taken in 2009.  We investigate the Fourier frequency dependent time lags in the light curves between the 0.3--1.0 keV and 2.0--5.0 keV energy bands as a function of source flux, including simultaneous modelling of the resulting lag-frequency spectra.  We find the shape of the lag-frequency spectra to vary significantly and systematically with source flux.  We model the lag-frequency spectra using simple transfer functions, and find that two time lag components are required, one in each energy band.  The simplest acceptable fits have only the relative contribution of the lagged component in the hard band varying with flux level, which can be associated with changes in the energy spectrum.  We discuss the interpretation of these results in terms of the currently popular models for X-ray time lags.
\end{abstract}



\maketitle
%

\section{Introduction}
\label{sect:intro}

Active galactic nuclei (AGN) are thought to be powered via accretion of gas onto a supermassive black hole (SMBH), with $\Mbh \sim10^{6}-10^{9} \Msun$, through an optically thick, geometrically thin disc (\citealt{shaksuny73}).  The release of gravitational potential energy in the disc manifests itself as thermal blackbody emission, peaking in the UV.  Some of these disc seed photons upscatter to X-ray energies through inverse-Comptonisation (\citealt{haardtmaraschi93}) in a corona of hot electrons.  While the X-ray emission mechanisms are well understood, many details about accretion flow and the geometry of the X-ray emission region still remain unexplained.

The leading model for X-ray variability of AGN (and X-ray binaries: XRBs) is based on the inward propagation of random accretion rate fluctuations in an accretion disc (e.g. \citealt{kotov01}, \citealt{king04}, \citealt{zdziarski05}; \citealt{arevalouttley06}, \citealt{ingramdone10}, \citealt{kelly11}), where the local mass accretion rate through the inner regions of the disc powers an extended X-ray emitting corona of hot electrons.  This model reproduces many of the currently known spectral variability patterns in AGN and XRBs (\citealt{arevalouttley06}).

The discovery of very short time lags ($\lesssim 100$\,s) between soft and hard X-ray bands (\citealt{fabian09}), where the soft band variations lag the hard band variations, may allow us to probe the X-ray emission region and inner accretion flow.  The soft lag is thought to represent the `reverberation' signal as the primary X-ray emission is reprocessed by relatively nearby material, one candidate for which is the inner accretion disc itself (e.g. \citealt{fabian09}).  These subtle signals directly encode the physical scales of the reflection region (by light travel time arguments).  The X-ray emission illuminates the optically thick disc material, producing the X-ray 'reflection' spectrum (\citealt{poundsetal90}; \citealt{georgefabian91}) which responds to variations in the X-ray luminosity with a short time lag set by the light travel time between the emission sites ($10 r_{\rmn{g}}$ is $\sim 100$\,s for $M_{\rm BH} \sim 2 \times 10^{6}$ \Msun, where $r_g = GM/c^{2}$).

Another model for the time lags has the X-rays scattering off a shell of circumnuclear material located at tens to thousands of $r_g$ from the central source of X-ray emission (\citeauthor{milleretal10a}, \citeyear{milleretal10b}; \citeyear{milleretal10a}, hereafter M10a and M10b respectively).  Electron scattering within this region produces a delayed, reprocessed X-ray continuum.  Absorption within the shell (bound-free and bound-bound transitions) means that, at large distances (hence longer time lags) soft X-rays are preferentially more likely to be absorbed than scattered out of the shell, relative to harder X-rays.  Including only the hard-band reprocessing, such a model can reproduce a (positive) hard lag at low frequencies, and if the edges of the shell are quite sharp it can also reproduce a (negative) soft lag in a narrow, higher frequency band, as a `ringing' artifact of the sharp impulse response function (a manifestation of the Gibbs phenomenon).  M10a extended the model to include reprocessing of the softer X-rays, from only inner regions of the shell, this reproduces a broader negative (soft) lag.

Soft lags have now been detected in many other sources (e.g. \citealt{demarco11}; \citealt{emmanoulopoulos11}; \citealt{zoghbi12a}; \citealt{cackett13}; \citealt{demarco13temp}) and even a source flux dependence on the soft lag has been seen (\citealt{kara13}).  This builds on the now well-established X-ray time lags observed at lower frequencies (e.g. \citealt{papadakis01}; \citealt{vaughan03a}; \citealt{mchardy04}; \citealt{arevalo06a}).  Soft lags have also been uncovered in black hole XRBs (\citealt{uttley2011}).  However, at present, the interpretation of the lags remains controversial (M10a; \citealt*{zoghbi11b}, hereafter Z11).

In this paper we investigate the reverberation signal in NGC 4051 with respect to source flux.  This is a narrow line Seyfert 1 galaxy that displays rapid and high amplitude X-ray spectral variability, including occasional prolonged periods of low flux and variability (e.g. \citealt{green99}; \citealt{uttley99}; \citealt{vaughan11a}).  \citet{mchardy04} revealed hard X-ray lags using a long \xmm\ observation taken in 2001, while \citet{demarco13temp} recently demonstrated the existence of soft X-ray lags at higher frequencies from more recent observations.  Modelling how the lag spectrum changes over different flux levels, with different spectral appearance, could provide valuable insight into the geometry of the reverberating medium, responsible for the soft band lags.

The rest of this paper is organised as follows.  Section \ref{sect:obs} discusses the observations of NGC 4051 and extraction of the X-ray light curves.  Section \ref{sect:lagfreq} describes the method of producing frequency domain products and modelling of the lag spectrum.  The implications of these results are discussed in section \ref{sect:disco}.  The results of simulations used to assess the reliability of the recovered lag-frequency spectrum are described in Appendix~\ref{ap:bias}.


\section{Observations and data reduction}
\label{sect:obs}

NGC 4051 was observed by \xmm\ $15$ times over $45$ days during May-June 2009 (see \citealt{alston13a} for observation details).  This paper uses data from the EPIC-pn camera \citep{struder01} only, due to its higher throughput and time resolution.  The raw data were processed from Observation Data Files (ODFs) following standard procedures using the \xmm\ Science Analysis System (SAS v12.0.1), using the conditions {\tt PATTERN} 0--4 and {\tt FLAG} = 0. The data reduction followed that of Vaughan et al. (2011) except that we used a 20 arcsec circular extraction region for the source.  The EPIC observations were made using small window (SW) mode; the fast CCD readout helps to mitigate event pile-up (\citealt{ballet99}; \citealt{davis01}).  We assessed the potential impact of pile-up using the SAS task {\tt  EPATPLOT}, and found only the highest flux revolution (rev 1730) showed significant signs of pile-up effects during the highest flux periods.  The data were filtered for high background flares, and the background was visually inspected for rises towards the end of each observation.  Strong background flares can introduce spurious time lags, so particular care has been taken to remove the influence of background variations (typically worse at the end of each observation).

The total duration of the useful data from the 2009 campaign is $\approx 572$ ks, giving $\sim 6 \times 10^{6}$ EPIC-pn source counts.  NGC 4051 was also observed for $\sim 120$\,ks in 2001 and $\sim 50$\,ks in 2002.  We do not include these data in the present analysis to avoid problems caused by systematic changes during the intervening $\sim 8$ years.


\section{Lag measurements}
\label{sect:lagfreq}

\begin{figure}
\includegraphics[width=0.44\textwidth]{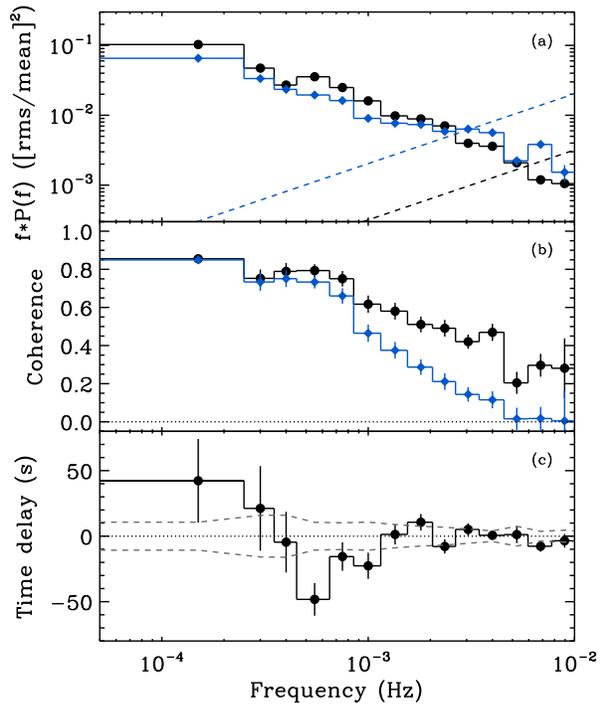}
\caption{Cross-spectral products for the soft (0.3--1.0 keV) and hard (2.0--5.0 keV) energy bands. Panel (a) shows power spectral density for the soft (black circles) and hard (blue diamonds) bands. The dashed lines are the Poisson noise estimates. Panel (b) shows the raw (blue diamonds) and Poisson noise corrected (black circles) coherence, see section~\ref{sect:lagfreq} for details. Panel (c) shows the time lag between the hard and soft band, where a positive values indicate the hard band lags. The grey dashed line is the Poisson-noise lag estimate, see section~\ref{sect:lagfreq} for details.}
\label{fig:lag_plot}
\end{figure}

\begin{figure*}
\begin{center}
\includegraphics[width=0.28\textwidth,angle=90]{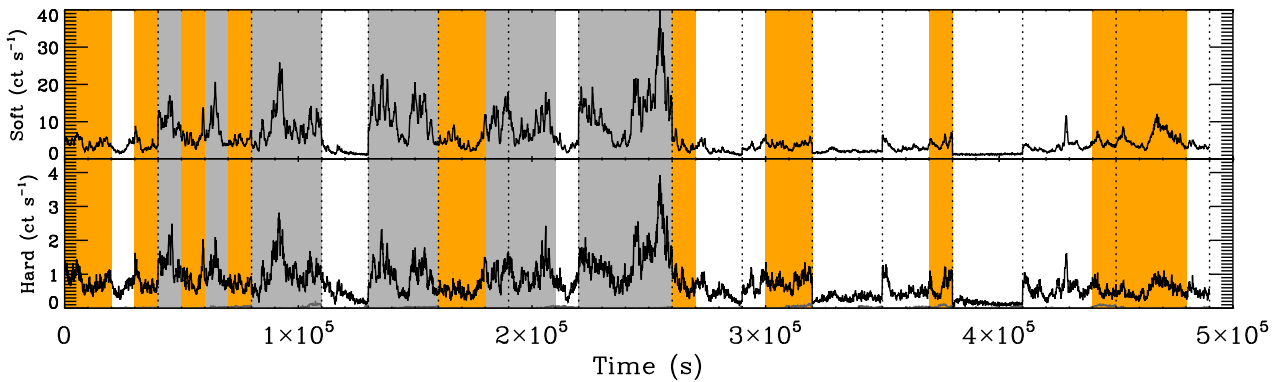}

\caption{Plot of soft (0.3--1.0 keV) and  and hard (2.0--5.0 keV) band light curve segments used in the fluxed analysis. The grey, orange and clear sections represent the high medium and low flux segments respectively, with a total band (0.2--10.0 keV) mean flux of 17.1, 7.8, and 4.4 ct s$^{-1}$. The vertical dotted lines separate the individual observations. The dark grey curve is the background level in each band.}

\label{fig:ltcrv}
\end{center}
\end{figure*}

From two evenly sampled time series $x(t), y(t)$ we can compute the Fourier Transforms $X(f), Y(f)$, in terms of amplitude phase.  We can estimate the power spectrum for each time series using $|X(f)|^{2}$ and $|Y(f)|^{2}$, after subtracting the Poisson noise and applying some normalisation factor.  Following e.g. \citet{vaughannowak97} we can define the complex valued cross-spectrum $C_{xy}(f) = X^{\ast}(f) Y(f) = \lvert X(f) \rvert \lvert Y(f) \rvert e^{i (\phi_{y(f)} - \phi_{x(f)})}$, where $\ast$ denotes the complex conjugate.  We estimated cross-spectral products by first averaging the complex $C_{xy}(f)$ values over non-overlapping segments of time series, and then averaging in geometrically spaced frequency bins (each bin spanning a factor $\sim 1.3$ in frequency).  For the analysis in this paper we use segment sizes of 10 ks and time bins of 10 s.  Accounting for observations that don't last an integer multiple of 10 ks, this leaves 490 ks of data from the 2009 observations.

From the resulting complex-valued cross-spectral estimate we obtain a phase lag $\phi(f) = \arg \langle C_{xy}(f) \rangle$ which may be transformed into the corresponding time lag $\tau(f) = \phi(f) / (2 \pi f)$.  This recovers the (time averaged) frequency dependent time lags between any correlated variations in $x(t)$ and $y(t)$.  Errors were estimated using standard formulae (\citealt{vaughannowak97}; \citealt{bendatpiersol86}).  The magnitude of the error estimates was checked using simulated light curves (see Appendix~\ref{ap:bias}).

From the cross-spectrum we can also obtain the coherence, a measure of the linear correlation between two time series as a function of Fourier frequency \citep{vaughannowak97}.  The coherence is defined as $\gamma_{xy}^{2}(f) = |\langle C_{xy}(f) \rangle|^{2} / \{\langle |X(f)|^{2} \rangle \langle |Y(f)|^{2} \rangle \}$, and takes on values in the range [0,1].  For time series contaminated by Poisson noise the coherence is suppressed, particularly at high frequencies where the intrinsic source variability is weaker and the Poisson noise tends to dominate.  The coherence can be corrected for the presence of Poisson noise as discussed in \citet{vaughannowak97}.

Fig.~\ref{fig:lag_plot} shows the power and cross-spectral products for the 0.3--1.0 keV (soft) and 2.0--5.0 keV (hard) bands, with mean count rates 5.8 and 0.8 $\rmn{ct\,s}^{-1}$ respectively.  The power spectral density (PSD) is plotted for each band in panel (a), where the hard band shows a flatter spectrum, as found in \citet{vaughan11a}.  Panel (b) shows the raw and noise corrected coherence.  The noise corrected coherence is high ($\sim 0.6$) for frequencies up to $\sim 10^{-3}$ Hz, above which it is seen to drop off smoothly.

Panel (c) in Fig.~\ref{fig:lag_plot} shows the time lag as a function of Fourier frequency (hereafter `lag-frequency spectrum').  Here we follow the convention of indicating a soft band lagging behind the hard band with a negative time lag (hereafter `soft lags').  The lag-frequency spectrum shape is similar to that seen in other sources, and seems to be quite common in low-redshift, X-ray variable AGN (e.g. \citealt{fabian09}; \citealt{emmanoulopoulos11} (hereafter E11); \citealt{demarco13temp}).  At frequencies less than $\approx 4 \times 10^{-4}$ Hz a hard lag is seen, whereas between $\approx 5 \times 10^{-4}$ and $1 \times 10^{-3}$ the soft emission lags the hard, with a maximum soft lag of $\sim 50$\,s at $\sim 6 \times 10^{-4}$\, Hz.

The effect of Poisson noise is to contribute an independent, random element to each phase difference measurement.  This makes small time lags difficult to detect at high frequencies where the Poisson noise begins to dominate over intrinsic source variations.  This effect can be estimated using equation 30 of \citet{vaughan03a}, which provides an estimate of the uncertainty on the time lag estimates due to the Poisson noise.  This is plotted in Fig.~\ref{fig:lag_plot} (panel c); we do not expect to recover reliable lags inside this range.  The PSD for the hard band was used in this calculation, as this has the lowest signal-to-noise of the two bands concerned.  For frequencies above $\sim 3 \times 10^{-3}$ Hz the Poisson noise dominates the lag.

The effects of spectral `leakage' (e.g. \citealt{vanderklis89}; \citealt{uttley02a}) on the lag-frequency spectrum were investigated using simulations (see Appendix~\ref{ap:bias}).  We found a bias in the size of the observed lag at a given frequency, such that the magnitude of the observed lag is reduced.  This implies that in the absence of such a bias the lags would be larger than currently estimated.


\subsection{Flux resolved time lags}
\label{sect:lagflux}

It is already known that the shape of the energy spectrum and the rms vary with flux level (e.g. Vaughan et al. 2011).  Here we investigate changes in the lag-frequency spectrum as a function of flux.  The time series segments were sorted by their average flux level and the cross-spectrum estimated in three flux bins, using an approximately equal number of light curve segments in each flux bin.  The total band (0.2--10 keV) mean count rate is 17.1, 7.8 and 4.5 ct s$^{-1}$ for the high, medium and low flux levels respectively, with 15, 15 and 19 segments (of 10 ks) in each flux bin, respectively.  The soft (0.3--1.0 keV) and hard (2.0--5.0 keV) band light curves are shown in Fig.~\ref{fig:ltcrv}, where the segments used in each flux level are indicated.  The segment length was chosen to be 10 ks as a compromise between having a wide range of fluxes (smaller duration segments give a wider range of mean fluxes) and having better low frequency coverage and reduced bias (longer segments give better low frequency coverage).  In Appendix~\ref{ap:bias} we explore the effect of using longer length segments.  The segment length is longer than the the timescale on which the PSD bends ($f \sim 2 \times 10^{-4}$\,Hz; \citealt{vaughan11a}) which also helps reduce the bias from very low frequencies.

\begin{figure}
\begin{center}
\includegraphics[width=0.48\textwidth,angle=0]{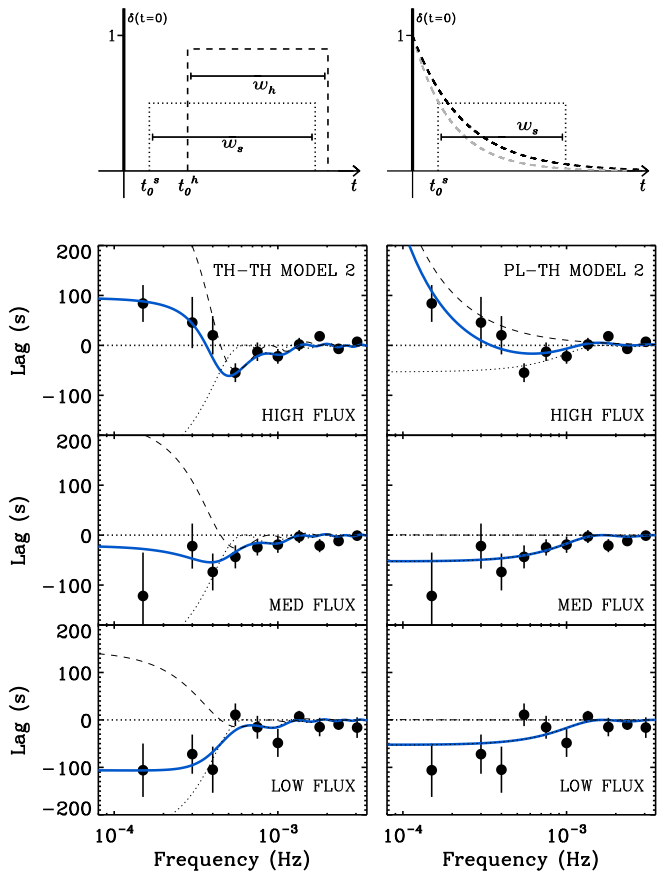}

\caption{Lag-frequency spectrum as a function of source flux between the soft (0.3--1.0 keV) and hard (2.0--5.0 keV) bands.  The data in the left and right hand panels are the same, with the panels from top to bottom showing the high, medium and low flux lag-frequency spectra respectively.  The solid blue lines are the results of fitting the response function models described in section~\ref{sect:lag_mod}.  Left: TH-TH model 2 of Table~\ref{tab:models} (upper section), consisting of a top hat in the hard band (dashed) and a top hat in the soft band (dotted).  Right: PL-TH model 2 of Table~\ref{tab:models} (lower section), consisting of a power law for the hard band (dashed) plus a top hat in the soft band (dotted).  The upper panels illustrate (since they are only schematic) the corresponding time response functions for the plotted models.  In the PL-TH model (right), the black and grey dashed lines represent the propagation of fluctuations between some initial fluctuations and the soft and hard emitting regions respectively.  The difference of the Fourier transforms of these two functions is approximately constant, giving lags between the two bands that have a power-law dependence on frequency.}

\label{fig:lag_flux_plot}
\end{center}
\end{figure}

Fig.~\ref{fig:lag_flux_plot} shows the lag-frequency spectrum, for the high, medium and low flux levels.  The high flux lag-frequency spectrum is consistent with that of the total flux shown in Fig.~\ref{fig:lag_plot}.  This is to be expected since the high flux intervals are also the most highly variable (Vaughan et al. 2011), so tend to have the higher Fourier amplitudes, and dominate the average cross-spectrum.  A clear change in the lag-frequency spectrum shape with source flux can be seen.

We ignore any frequency bins where the raw coherence is consistent with zero (see Fig.~\ref{fig:lag_plot}).  This leaves 10 frequency bins per lag-frequency spectrum after applying the frequency averaging.  The noise corrected coherence is high ($\gtrsim 0.6$) for these 10 frequency bins in all three flux ranges.  No significant changes in the coherence spectrum were seen across the flux levels.


\subsection{Modelling the lag}
\label{sect:lag_mod}

In this section we attempt to fit the lag spectra for the high, medium and low fluxes using simple analytical models.  We assume the time series obtained from each band is actually some underlying `source' variability $z(t)$ convolved with a linear impulse response function\footnote{We use `impulse response' or `response function' to indicate the function as applied in the time domain (by convolution), and `transfer function' to represent its Fourier counterpart, as these terms are standard in the signal processing literature.}: $x(t)=\psi_x(t) \otimes z(t)$ and $y(t)=\psi_y(t) \otimes z(t)$.  This is clearly a simplification but allows us to test some `toy models' for the origin of the time lags\footnote{We note that the response function models discussed here or elsewhere in the literature are linear, which give unity coherence between the input and output time series, and therefore fail to reproduce the coherence spectrum (see Fig.~\ref{fig:lag_plot}).}.  The phases obtained from the cross-spectrum $C_{xy}(f)$, which give the time lag-frequency spectrum, are (on average) the differences between the phases of the Fourier transforms of each response, i.e. $\phi(f) = \arg \Psi_y(f) - \arg \Psi_x(f)$.

Previous work (e.g. M10a, E11, Z11) has made use of two types of time lag model: a power law dependence of the time lag with frequency, or the delay due to a top hat impulse response.  A roughly power law (PL) frequency dependence of the hard X-ray time lag is well established in X-ray binaries (e.g. \citealt{miyamoto88}; \citealt{nowak99}; \citealt{pottschmidt00}), at least at low frequencies, and consistent results from AGN have been reported several times (e.g. \citealt{papadakis01}; \citealt{vaughan03a}; \citealt{mchardy04}; \citealt{arevalo06a}).  We therefore consider a PL model of the form $\tau(f) = Nf^{-\alpha}$ with $\alpha \approx 1$.  The time domain representation of this function is shown in Fig.~\ref{fig:lag_flux_plot}.  This may represent the effects of variations propagating through and modulating an extended emission region \citep[e.g.][]{kotov01, arevalouttley06}.  But a PL lag-frequency spectrum does not show a reversal in the direction of the lag (as seen in the data), meaning alternative or additional terms are required to explain the negative lag.

A simple alternative model is provided by a `top hat' (TH) impulse response, as might be produced by reverberation from an extended medium.  The TH is characterised by three parameters: start time $t_{0}$, width $w$, and scaling fraction $S$.  In practice we fit the time-lag spectrum using the time lags extracted from the Fourier transform of a TH plus a $\delta$-function.  The $\delta$-function corresponds to direct (not delayed or smoothed) emission, and the parameter $S$ sets the intensity of delayed emission relative to the direct emission.  The start time ($t_{0}$) of the top hat component was allowed to be negative to allow for time lags of the direct component, as is found in \citet{wilkins12}.  As discussed in M10a and Z11, a TH response function applied to the hard band alone can produce both a hard lag at low frequencies and an apparent soft lag at higher frequencies due to `ringing' caused by the sharp edges of the response function in the time domain.  However, as pointed out by Z11, such a model predicts a frequency range for the soft lag that is narrower than typically observed.

We fitted various combinations of PL and TH-based time lag models to our data.  As the time lags were estimated by averaging the cross-periodograms from multiple segments their distributions should be approximately Normal, and hence we can use standard $\min(\chi^2)$ fitting techniques.  The lag-frequency spectra from the three flux levels were fitted simultaneously.  Some parameters were `tied' such that the same value applies to the model at all three flux levels, and other parameters were `untied' meaning that the parameter was allowed to take on different values for each flux level.  During the fitting we experimented with different combinations of tied and untied parameters, as summarised in Table~\ref{tab:models}.

As expected, the simple PL model and the simple hard-band TH model do not provide acceptable fits to our data (with $p < 10^{-4}$ in a $\chi^2$ goodness-of-fit test).  We therefore consider in detail two more complex models.  The first is based on TH plus $\delta$-function responses in the soft and hard bands, the second model adds lags from a TH plus $\delta$-function response in soft band to a PL lag in the hard band.  We refer to these as the TH-TH and TH-PL models, respectively (they correspond to models 2 and 3 of E11).  A model with PL-like responses in both bands (PL-PL model) can in principle generate hard lags on long timescales and soft lags on short timescales, if the PL index in the hard band is steeper than the soft band.  We did not attempt to fit such a model as it necessarily gives only a very smooth lag-frequency spectrum, which is inconsistent with the relatively sharp transition from hard to soft lags observed.

\begin{table}
 \caption{Results of simultaneous model fitting to the three lag spectra, for variants of tied (T) and untied (U) parameters. Column (1) is the model variant, columns (2-7) are the model parameters, column (8) is the ${\Large \chi}^{2}$ fit and degrees of freedom, and column (9) is the null hypothesis probability.  The $h,s$ sub/super-scripts indicate the parameter is for the transfer function in the hard and soft band respectively.}

 \begin{tabular}{@{}lcccccccc}
 \hline
Variant	& \multicolumn{6}{c}{Parameters: Tied / Untied} & ${\Large \chi}^{2}$ / dof& p value \\
 \hline
 \multicolumn{9}{c}{TH-TH Model} \\
	& $t_{0}^{\rmn{h}}$ 	& $w_{\rmn{h}}$	& $S_{\rmn{h}}$	&$t_{0}^{\rmn{s}}$ 	& $w_{\rmn{s}}$	& $S_{\rmn{s}}$	&	& \\
1	& T 	& T	& T	& T	& T	& T	& 39/24 & 0.02\\
2	& T 	& T	& U	& T	& T	& T	& 19/22 & 0.66\\
3	& T 	& T	& T	& T	& T	& U	& 21/22 & 0.52\\
4	& T 	& T	& U	& T	& T	& U	& 17/20 & 0.60\\
5	& U 	& T	& T	& U	& T	& T	& 24/20	& 0.24\\
6	& U 	& U	& T	& U	& U	& T	& 14.5/16 & 0.56\\
 \hline 
 \multicolumn{9}{c}{PL-TH Model} \\
	& $N_{\rmn{h}}$ 	& $\alpha_{\rmn{h}}$ &$t_{0}^{\rmn{s}}$ 	& $w_{\rmn{s}}$	& $S_{\rmn{s}}$&	&	& \\
1	& T 	& T	& T	& T	& T	& 	& 39/25 	& 0.04\\
2	& U 	& T	& T	& T	& T	& 	& 25.6/23 	& 0.32\\
3	& T 	& T	& T	& T	& U	& 	& 35/23 	& 0.05\\
4	& U 	& T	& T	& T	& U	& 	& 22.6/21 	& 0.37\\
5	& U 	& U	& T	& T	& U	& 	& 22.5/19	& 0.22\\
6	& T 	& T	& U	& U	& U	& 	& 34/19 	& 0.02\\
7	& T 	& U	& U	& U	& U	& 	& 24/17 	& 0.14\\
 \hline
\end{tabular}
\label{tab:models}
\end{table}

\begin{table}
 \caption{Parameter values for the simplest model fits to the three flux lag-frequency spectra.  These correspond to the TH-TH variants 2 and 3, and PL-TH variants 2 and 3 from Table 1.   Column (1) is the model parameter, where the $h$ and $s$ sub/super-scripts indicate the parameter is for the transfer function in the hard and soft band respectively.  Columns (2-4) are the tied or untied parameter values, where `-' means the parameter was tied for all fluxes (see section~\ref{sect:lag_mod} and Table~\ref{tab:models} for the model variant details).}
 \label{tab:model_best}
 \begin{tabular}{@{}lccc}
 \hline
Parameter		&High / all		&Medium		&Low \\
 \hline
 \multicolumn{4}{c}{TH-TH variant 2} \\
$t_{0}^{\rmn{h}}$	&$1120 \pm 120$		& -		& -\\
$w_{\rmn{h}}$		&$560 \pm 60$		& -		& - \\
$S_{\rmn{h}}$		&$0.45 \pm 0.09$ 	&$0.26 \pm 0.08$&$0.14 \pm 0.07$ \\
$t_{0}^{\rmn{s}}$	&$60 \pm 180$		& -		& - \\
$w_{\rmn{s}}$		&$1530 \pm 230$		& -		& - \\
$S_{\rmn{s}}$		&$0.44 \pm 0.09$	& -		& - \\

 \multicolumn{4}{c}{TH-TH variant 3} \\
$t_{0}^{\rmn{h}}$	&$1130 \pm 100$		& -		& -\\
$w_{\rmn{h}}$		&$570 \pm 70$		& -		& - \\
$S_{\rmn{h}}$		&$0.28 \pm 0.07$ 	& -		& - \\
$t_{0}^{\rmn{s}}$	&$-10 \pm 110$		& -		& - \\
$w_{\rmn{s}}$		&$1630 \pm 130$		& -		& - \\
$S_{\rmn{s}}$		&$0.23 \pm 0.10$	& $0.53 \pm 0.13$&$0.66 \pm 0.15$ \\
 \hline
 \multicolumn{4}{c}{PL-TH variant 2} \\
$N_{\rmn{h}}$		&$0.0005\pm0.0001$	&$0.0\pm0.0002$		&$0.0\pm0.0002$	\\	
$\alpha_{\rmn{h}}$	&$1.44\pm0.07$	& -		& -		\\
$t_{0}^{\rmn{s}}$ 	&$-10\pm110$	& -		& -		\\
$w_{\rmn{s}}$		&$610\pm120$	& -		& -		\\
$S_{\rmn{s}}$		&$0.22\pm0.05$	& -		& -		\\

 \multicolumn{4}{c}{PL-TH variant 3} \\
$N_{\rmn{h}}$		&$0.0002\pm0.002$&-		& -		\\	
$\alpha_{\rmn{h}}$	&$1.42\pm0.10$	& -		& -		\\
$t_{0}^{\rmn{s}}$ 	&$10\pm80$	& -		& -		\\
$w_{\rmn{s}}$		&$610\pm100$	& -		& -		\\
$S_{\rmn{s}}$		&$0.16\pm0.11$	&$0.29\pm0.12$	&$0.25\pm0.12$	\\

 \hline
 \end{tabular}
\end{table}

Ignoring flux-dependence, the TH-TH model has six parameters (three for each TH component). Table~\ref{tab:models} shows the results of fitting the model with variants of tied and untied parameters.  The simplest acceptable fit to the data was provided by letting only the scaling factor $S_h$ vary with flux (this is listed as TH-TH $2$ in Table~\ref{tab:models}; $8$ free parameters). In this case a good fit is obtained if the $S_h$ parameter, indicating the strength of the delayed component in the hard band, decreases with decreasing flux and is consistent with there being no delayed component at low fluxes.  This is to be expected given the disappearance of the hard lag at low frequencies in the lower flux data (Fig 3; left panels).  Allowing only the $S_s$ parameter to vary (model TH-TH $3$ in Table~\ref{tab:models} gives a worse, but still acceptable fit to the data.  Keeping the $S_s$ and $S_h$ constant with flux, but allowing the start times ($t_s$ and $t_h$) to vary with flux provided a rather worse, but still acceptable, fit to the data (model TH-TH $5$ in Table~\ref{tab:models}).  Allowing the top hat widths ($w_s$ and $w_h$) to also vary independently with flux gave a very good fit (model TH-TH $6$ in Table~\ref{tab:models}) but with the largest number of free parameters ($14$).  The fit parameter values for TH-TH models 2 and 3 are listed in Table~\ref{tab:model_best}.

Again ignoring flux-dependence the PL-TH model has five parameters: $N_h$,$\alpha_{h}$, $t_{0}^{s}$, $w_{s}$, $S_{s}$.  The results of fitting this model to the three lag-frequency spectra are displayed in Table~\ref{tab:models}.  As with the previous model, the simplest variant providing an acceptable fit allowed only the hard band scaling factor $N_h$ to vary with flux (PL-TH model $2$ in Table~\ref{tab:models}).  In particular, the parameter $N_h$, representing the normalisation of the PL lag in the hard band, is positive for the high flux data and consistent with zero for medium and low flux data, which accounts for the disappearance of the hard lag at low frequencies in the lower flux data (Fig 3; right panels).  Allowing only $S_s$ to vary with flux (PL-TH model $3$ in Table~\ref{tab:models} did not give an acceptable fit.  The fit parameter values for PL-TH models 2 and 3 are listed in Table~\ref{tab:model_best}.  

We have also tried a model combining features of both the above models, namely TH plus $\delta$-function applied to the soft band, and a TH plus $\delta$-function applied to the hard band, with additional hard lags from a PL component.  In this case the power-law normalisation is always consistent with zero (for various combinations of tied/untied parameters), and so the model reduces to the TH-TH model above.

The models were refitted after re-parameterising the scaling factors such that the TH component normalised to unity and the normalisation of the delta function was a free parameter.  This is mathematically equivalent, but allows us to monitor any changes in the direct component with flux.  The results are consistent with our model fits in Table 1, where we found the scaling fraction on the $\delta$-function in the soft band to be constant with flux.

An extreme variation on this model has direct component in the soft band vanishing at low fluxes. We explicitly tested this by fixing the $\delta$-function scaling factor in the soft band at low flux to zero.  For the TH-TH and PL-TH model variants described in Table 1, we again simultaneously fit across all flux levels.  We find fixing the $\delta$-function scaling factor to zero did not provide an acceptable fit ($p < 0.01$ for all model variants).


\section{Discussion and conclusions}
\label{sect:disco}

\subsection{Summary and comparison with previous work}

We have studied the frequency dependent time delays in the Seyfert galaxy NGC 4051 using a series of \xmm\ observations made in 2009.  We found the lag-frequency spectrum varies significantly and systematically with source flux.  At high fluxes there is a hard (`positive') lag at the lowest frequencies and a soft (`negative') lag confined to a narrow range of frequencies around $\approx 5 \times 10^{-4}$ Hz, but at lower fluxes the low frequency lag reverses direction, becoming a soft lag.  The lag-frequency spectra in three flux bins can be modelled simultaneously using simple response function models.  The simplest acceptable fits have all parameters tied between the flux levels, except for the parameter that determines the magnitude of the delayed component in the hard band light curve.  Allowing just the scaling fraction in the soft band to vary provides an acceptable fit in the TH-TH model, but not in the PL-TH model.

Previous work by \cite{mchardy04}, M10b and \cite{legg12} on X-ray time lags in NGC 4051 found hard lags increasing to lower frequency, but did not identify a soft (`negative') lag. This may reflect genuine differences in the variability between observations, but might also be due to differences in the data quality and/or analysis methods that meant a soft lag went undetected.  We note that these papers all used data obtained during times when NGC 4051 was relatively bright, and our analysis shows that at high fluxes the soft lag occurs over a relatively narrow range of frequencies.  M10b and Legg et al. (2012) used data from \emph{Suzaku} which contain periodic gaps (due to the $\sim 96$ min Low Earth orbit), and used a much longer time bins (256\,s) than our analysis.  It may well be the case that the data or analysis used by these authors are less sensitive to the short ($\sim 50$\,s) lag we observe with \emph{XMM-Newton}.  Recently, \cite{demarco13temp} published a lag-frequency spectrum for a sample of unabsorbed, radio-quiet AGN including NGC 4051.  Their lag-frequency spectrum represents an average over \xmm\ observations taken during 2001, 2002 and 2009, and is consistent with the lag-frequency spectrum we show for high flux periods (which will tend to dominate such an average, as discussed above).

\subsection{Understanding the time delays}

The changes in lag-frequency behaviour with flux could be linked to the strong change in spectral shape with flux (see e.g. Vaughan et al. 2011) if the hard and soft lag components are linked to the components which make up the energy spectrum, and these components change in relative strength.  A detailed comparison of the energy-dependent behaviour is beyond the scope of this paper, however.  But we note here that the lack of positive lags at low fluxes could indicate a change in the circum-nuclear structure, and also be responsible for the dramatic change in the energy spectrum.

For the purpose of the present discussion we assume the cause of the very lowest fluxes is no different from the X-ray variability in general, simply the lowest point in a continuum of spectral variability.  This seems most natural given the rms-flux relation for NGC 4051 (\citealt{uttleymchardyvaughan05}; Vaughan et al. 2011) appears as a single linear track from highest to lowest fluxes, and the PSD shape remains constant in shape for high and low fluxes (Vaughan et al. 2011)\footnote{The flux-resolved PSD analysis of Vaughan et al. (2011) was carried out using a broad (0.3-10 keV) energy band. But, as the count spectrum of NGC 4051 is dominated by soft (\textless 2 keV) photons, this is effectively tracks the soft X-ray PSD with flux, which should be most sensitive to changes in absorption.}.  If the soft band emission was dominated by reprocessed emission at low fluxes we would expect to see a change in the PSD (the reprocessor acts as a low-pass filter).  The optical and X-ray variations (including low flux periods) are correlated on timescales of $\gsim 1$ day (\citealt{breedt10}), and the UV and X-ray variations are correlated on timescales of $\sim 1$ hr (\citealt{alston13a}).  These correlation results are independent of X-ray flux, and support the model where X-ray variability on the timescales that soft lags are observed is dominated by intrinsic X-ray luminosity variations.

We fitted the lag-frequency spectrum using two simple kinds of impulse response models: a top hat (TH) response in the time domain and a time lag that varies with frequency as a power law (PL).  The soft  band lag of about $\sim 50-100$\,s can be explained by a single TH response function spanning lags from $0$\,s to $\sim 10^3$\,s.  The observed time lag is diluted by the direct (zero delay) component included in the response function.  The hard band lag can be explained either in terms of a top hat response with a minimum lag of $\sim 10^3$\,s, or a power law-like frequency dependence with an index of $\alpha \approx 1.5$.  These two models give similar quality fits over the observed frequency range, but diverge strongly at lower frequencies where these observations provide little sensitivity.

If the time lags from the response function models are interpreted simply as the result of light travel time differences, the maximum time lag in each band gives a distance to the most distant parts of the reprocessor of $\tau \sim (1 - \cos \theta) R/c$ where $R$ is the distance between source and reprocessing region, and $\theta$ is the angle between source-reprocessor line and the line-of-sight to the source.  Assuming $\theta > \pi/2$ (i.e. the reprocessor is not concentrated entirely on the near side of the source), the maximum lag corresponds to $\sim R/c$.  The best fitting TH-TH models have the maximum time lag in the hard and soft band consistent with each other at a value $\sim 1600$\,s, placing the outer region of the reprocessor at $\sim 160 r_{g}$ for NGC 4051 (assuming $M_{\rm BH} \sim 2 \times 10^{6}$ \Msun; \citealt{denney09}).  In the best fitting PL-TH model the soft band response extends out to $\sim 600$\,s, placing the reprocessor at $\sim 60 r_{g}$.

Models using a single delayed component, i.e. contributing to either the hard or soft band but not both, give unacceptable fits to the lag-frequency spectrum.  This, combined with the apparent flux-dependence of the contribution of the hard lag but not soft lag components, supports the need for at least two components causing time lags, one affecting each band.  This is consistent with a more complex model in which time lags are caused by a combination of propagating fluctuations and reflection (e.g. E11; Z11).  It could also be the case that a single lagging component contributes to both bands, but its spectral shape changes with flux (so that it appears as a constant fraction of variable emission in the soft band but decreasing with flux in the hard band).

The apparent change in direction of the lag at low fluxes can be explained by a change in the responses: as the flux decreases the skew towards lagging values becomes weaker in the hard band, relative to that in the soft band.  This could occur either if the hard lagging component gets weaker compared to the hard direct light, or if the soft lagging component gets stronger relative to the soft direct light.  If the direct emission, with a power law energy spectrum, contains `intrinsic' hard time delays, as supposed in e.g. the propagating fluctuations model (Kotov et al. 2001; Ar\'{e}valo \& Uttley 2006) there will be hard delays even in the absence of reprocessing. The `external' reprocessing then adds a second source of delays, contributing a short lag extending up to higher frequencies. In this scheme the efficiency of the `external' reprocessing remains constant with flux, but the fraction of delayed light making up the intrinsic emission must decrease with flux.


If the lower fluxes are due to a low intrinsic (isotropic) luminosity, and the properties of the reprocessing medium remain constant, then the time lags should be relatively constant with flux.  On the other hand if the low flux is a result of increasing absorption along the line-of-sight (with a relatively constant isotropic luminosity) then the contribution of the delayed components in the light curves should increase (relative to the direct components) as the flux decreases.  This is one way to interpret the TH-TH 3 model.

An extreme variation on this idea is that the periods of low flux and variability are the result of absorption completely obscuring the direct component in the soft band, but not the (more extended) reprocessing region.  The low flux lag spectrum may well be reproduced by the lack of a $\delta$-function at $t=0$ in the soft band response function.  However, we found this model, in which the direct (zero delay) contribution to the soft band disappears at low fluxes, did not give a good fit to the lag-frequency data.

M10a proposed to explain both hard and soft time lags in terms of scattering of X-rays in a spherical shell of material located at tens to thousands of $r_g$ from the central source.  If the reprocessing in each band is the result of scattering within the shell, and hard X-rays penetrate deeper into the shell as suggested by M10a, then the maximum lag in the response function due to the reprocessed component should be larger for the harder band, while the minimum lag should be the same in the two bands.  Our best fitting TH-TH models (variants 2 and 3) are at odds with these predictions; the best fits have similar maximum lag in the two bands, but a larger minimum lag in the hard band.

\section*{Acknowledgements}

We thank the referee for providing a thorough and helpful report on our paper which helped us to improve the original manuscript.  WNA acknowledges support from an STFC studentship.  This paper is based on observations obtained with \xmm, an ESA science mission with instruments and contributions directly funded by ESA Member States and the USA (NASA).  This research has made use of NASA's Astrophysics Data System Bibliographic Services.


\bibliographystyle{mn2e}
\bibliography{4051_lags}


\label{lastpage}

\appendix

\section{Bias in time lag estimation}
\label{ap:bias}

The X-ray PSD of NGC 4051 shows substantial power, and is still rising to lower frequencies, below $10^{-4}$ Hz, the lowest frequency used in our cross-spectrum analysis (due to the choice of $10$ ks segments).  One side-effect of observing a low-frequency dominated noise process using short segments is known as `spectral leakage' -- a bias on the Fourier transform caused by the side-lobes of the Fourier transform of the window function.  This is known to distort PSD estimates (\citealt{vanderklis89}; \citealt{uttley02a}; Vaughan et al. 2003), but similarly distorts the phase estimates, and hence the cross-spectrum of two time series.  Here, this in appendix we briefly explore the origin and consequences of this `leakage' on the lag-frequency spectrum.

\subsection{Origin of the phase bias}
\label{ap:origin}

It is possible to understand the origin of the phase (time) lag bias in 
terms of the finite segment length, or {\it sampling window}, used to 
make each cross-spectrum estimate. The following analysis sketches out 
the origin of the bias on the phase lag between two time series $x(t)$ 
and $y(t)$.

If $X(f)$ and $Y(f)$ are the Fourier Transforms of $x(t)$ and $y(t)$, 
then the complex-valued cross-spectrum is formed from the ensemble 
average of their product:
\begin{equation}
   \langle C(f) \rangle = \langle X^{\ast}(f) Y(f) \rangle
\end{equation}
and the phase lag-frequency spectrum $\phi(f)$ is obtained from its argument:
\begin{equation}
  \phi(f) = \arg \langle C(f) \rangle = \arctan \left( \frac{\langle 
q(f) \rangle}{\langle c(f) \rangle} \right)
\end{equation}
where $q(f)$ and $c(f)$ are the estimated `quadrature-spectrum' and 
`co-spectrum', the imaginary and real components of $C$, respectively. 
These are each real-valued and may be computed from:
\begin{eqnarray}
\label{eqn:quad-co}
   q(f) =  \Re(X)\Im(Y) -  \Im(X)\Re(Y)  \nonumber \\
   c(f) =  \Re(X)\Re(Y) +  \Im(X)\Im(Y)
\end{eqnarray}
where $\Re(\cdot)$ and $\Im(\cdot)$ indicate the real and imaginary 
components of some variable.

Real time series have finite length and sampling, $x_n(t_i)$ where 
$i=1,2,\ldots,n$ and $t_i = i\Delta t$. We can understand the effect of 
finite length time series by treating the observed time series 
$x_n(t_i)$ as the product of an infinitely long series $x(t)$ with a 
window function $w(t)$ that is zero everywhere except $-T/2 \le t \le 
T/2$, where $w(t)=1$ and $T = n \Delta t$. (We shall neglect the effect 
of the finite time resolution $\Delta t$, but the following results are 
approximately correct for small $\Delta t$.)
\begin{equation}
   x_n(t)  =  x(t)w(t)
\end{equation}
and by the convolution theorem this leads to the relationship between 
the observed and asymptotic Fourier components:
\begin{equation}
\label{eqn:finite-ft}
    X_n(f_j)  =  \int_{-T/2}^{T/2} X(f) W(f - f^{\prime}) d f^{\prime}
\end{equation}
at Fourier frequencies $f_j = j/ T$ ($j=1,2,\ldots,n/2$). Here, $W(f)$ 
is the Fourier transform of the function $w(t)$, with the form $W(f) = 
\sinc(\pi T f)$ for a rectangular window. More generally, for discretely 
sampled series this is the \emph{Dirichlet kernel}. See \cite{priestley81} 
and \cite{jenkinswatts69} for detailed discussion of the effect of 
window functions on Fourier products such as auto- and cross-spectral 
densities. This convolution of $X(f)$ with $W(f)$ distorts the real and 
imaginary components of $X_n(f_j)$, and likewise for $Y_n(f_j)$. In 
practice, when we estimate the phase lags using $x_n(t)$ and $y_n(t)$ we 
replace the Fourier transforms in equations \ref{eqn:quad-co} with those 
of equation \ref{eqn:finite-ft}. The distortion of each of the 
components -- $\Re(X_j)$, $\Im(X_j)$, $\Re(Y_j)$, $\Im(Y_j)$ -- 
generally leads to a bias on the estimated co- and quadrature spectra. 
The consequence is that if the phase difference $\phi(f)$ changes with 
$f$ the distortion on co- and quadrature spectrum will be different, and 
lead to a bias on the phase lag $\phi_n(f_j)$ calculated from their 
ratio. The bias of the Fourier transforms, and hence the phase 
difference between them, decreases as $W(f) \rightarrow \delta(f)$, i.e. 
as $T \rightarrow \infty$.

If one or more of the PSDs are rising steeply to low frequencies, this 
can lead to a `leakage' of the Fourier components from low (unobserved) 
to higher (observed) frequencies, through the side-lobes of the kernel. 
But even if the PSDs are flat towards lower frequencies, Fourier 
components at nearby frequencies influence $X_n(f_j)$ and $Y_n(f_j)$ and 
so, even in the absence of `red noise leakage' of the power density 
there can be a non-trivial bias on the phase lags. This is strongest 
when the phase spectrum $\phi(f)$ changes rapidly with $f$, i.e. has 
large curvature $|d^2\phi(f)/df^2|$. By contrast, when the phase 
spectrum is very smooth (approximately constant or linear over 
$\sim$few$\Delta f$, where $\Delta f = 1/n \Delta t$) the phases at 
nearby frequencies are sufficiently similar that the distortion on the 
co- and quadrature spectrum is similar and cancels when computing the 
phase lag, leading to small bias.
\subsection{Simulations of leakage}
\label{ap:sim}

\begin{figure}
\begin{center}
\includegraphics[width=0.45\textwidth,angle=0]{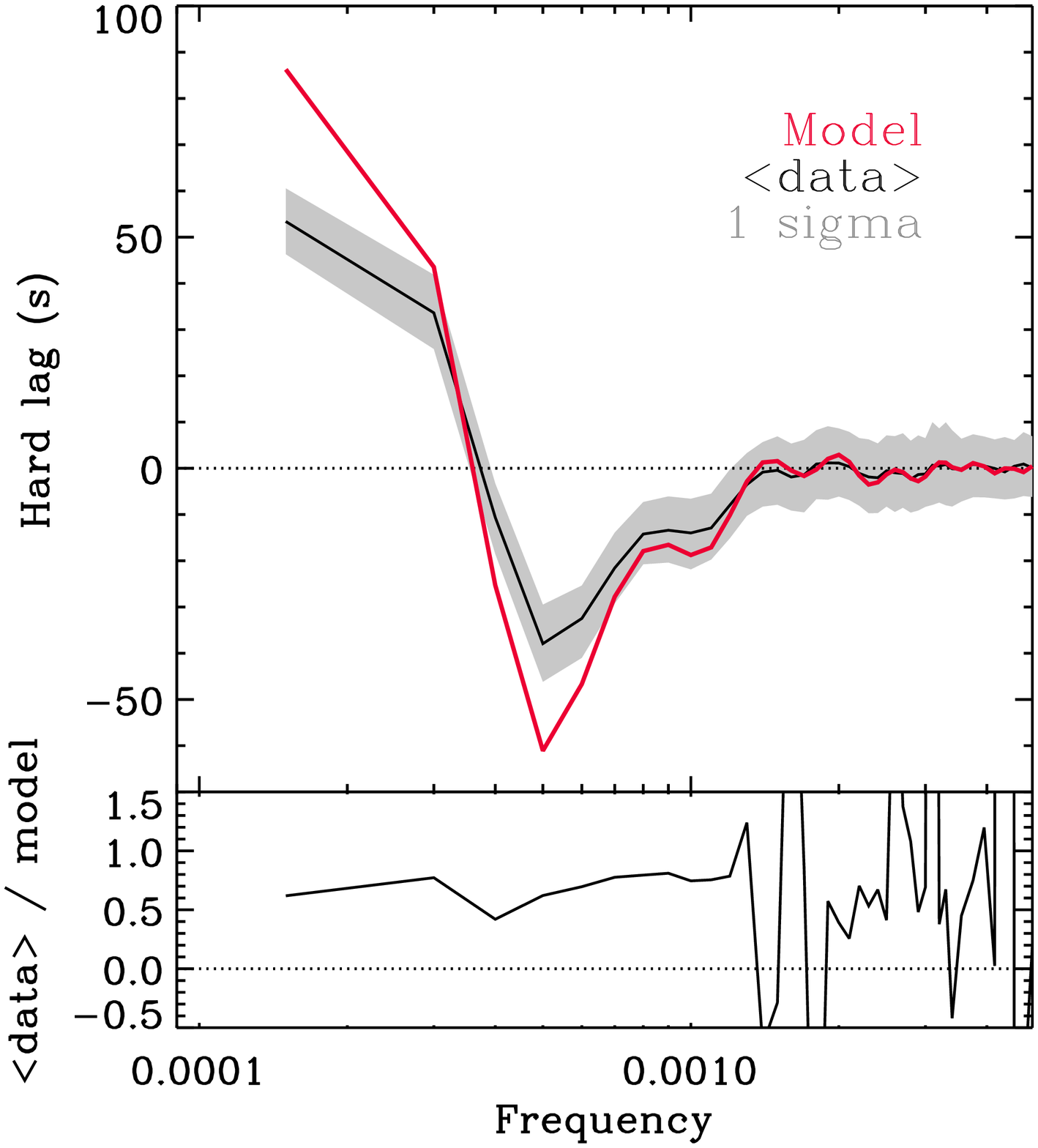}

\caption{The effect of leakage on the measured lag-frequency spectrum.  The upper panel shows the lag-frequency spectrum for simulated pairs of light curves with a frequency-dependent time lag.  The red curve is the model time lag, the black line is the mean measured lag and the grey shaded region is the 1 sigma confidence intervals from the simulations.  The model shown is the best fit TH-TH (see section~\ref{sect:lag_mod}).  The lower panel shows the ratio between the mean measured lag and the model.}

\label{fig:leakage}
\end{center}
\end{figure}

To assess the level of phase bias in the observed lag-frequency spectra we simulated $1000$ pairs of random time series.  For each pair we used the \citet{timmerkonig95} method to generate a pseudo-random, Gaussian time series of length 50 days and 10\,s binsize, corresponding to much longer and shorter timescales than is measured in the lag-frequency spectrum. The PSD used to generate the time series was a bending power-law with low frequency slope $-1.1$, high frequency slope $-2.0$ and break frequency $\nu_{\rmn{B}} = 2 \times 10^{-4}$ Hz \citep{vaughan11a}. For each of these simulations a second time series was produced using the same Fourier amplitudes and phases, except for an additional frequency-dependent phase shift derived from the time lag models described in section~\ref{sect:lag_mod}. For each pair of time series, $15$ intervals were extracted using the sampling pattern as the real \emph{XMM-Newton} observations. The rms-flux relation (see \citealt{uttleymchardyvaughan05}) was added to the simulated time series using the static exponential transformation, although we note the results are largely insensitive to this. Observational noise was added to each time series by drawing a Poisson random deviate with mean equal to the mean count-per-bin in the real light curves.

Fig.~\ref{fig:leakage} shows the input model of the time delays (TH-TH model in this case) compared to the average lag-frequency spectrum from the simulated data, clearly revealing a bias in the estimates. The effect of leakage is to suppress the magnitude of the observed time lag. The ratio plot in Fig.~\ref{fig:leakage} shows there is a $\sim 30\%$ reduction in the recovered lag for frequencies $\lesssim 2 \times 10^{-3}$\,Hz.  One implication is that the measured time lags would be larger if the bias towards zero could be removed.  We find that changing the top-hat width or start-time ($w$; $t_{0}$) has little effect on the magnitude of the bias. A slightly smaller bias towards zero lag is observed in simulations with the PL-TH model.

\subsection{Covariance between lag estimates}
\label{ap:covar}

Another consequence of leakage is covariance between lag estimates in adjacent frequencies \citep[chap. 9]{jenkinswatts69}.  The simulations performed here were used to constrain the degree of covariance.   In the limit of very long observations (and in the absence of biases due to spectral leakage) the covariance between lag estimates at different frequencies tends to zero, but for finite length time series there will remain some small covariance.  From the covariance matrix of the simulated time lags as a function of frequency, the correlation between adjacent frequency bins can be calculated.  The lowest frequency bins show the highest correlation with each other (but with values $\lesssim 0.25$).  A histogram of the correlation between adjacent frequencies shows a Gaussian distribution centred on $\sim$zero.  This shows that, whilst some degree of covariance exists between adjacent frequencies, treating the lag measurements at each frequency as independent should not substantially distort the results of any $\min(\chi^2)$ fitting techniques.  The scatter in the simulated lags was used to check the magnitude of the standard error estimates (e.g. \citealt{vaughannowak97}; \citealt{bendatpiersol86}), which show consistent results.

\subsection{Reducing leakage bias}
\label{ap:reduce}

We have explored variations of the lag estimation procedure to reduce the bias caused by spectral leakage.  We computed the lag-frequency spectrum using 20\,ks segments (not shown here).  Again the mean source rate in each segment was chosen to give an approximately equal number of segments for 3 flux levels.  The segments used therefore differ from those shown in Fig.~\ref{fig:ltcrv}, and the range of mean fluxes is reduced compared to the shorter segments used in the above analysis.  However, the same change in the lag-frequency spectrum is observed with source flux.  We have also computed the lag-frequency spectrum using whole observations ($25-40$\,ks) as the segment length, before binning in frequency space over segments of differing length.  This includes more data in the lag-frequency estimate, but gives a much smaller range of mean fluxes.  We can therefore only compute this for two flux levels, which still show the same change in the lag-frequency spectrum with source flux.  We conclude that the systematic change in the lag-frequency spectrum with source flux is robust to the details of the analysis.

Leakage bias can be reduced by `end-matching' the data, whereby a linear trend is removed such that the first and last points are level (\citealt{fougere1985}).  This `end-matching' removes, to a large extent, linear trends from the data, and alleviates the problem caused by circularity of the Fourier transform when estimating the cross-spectrum.  It removes the bias coming from lower frequencies ``leaking" in, but does not remove the bias from nearby frequencies.  We computed the lag-frequency spectrum for the three flux levels using 10 ks segments, where each segment is end-matched individually.  The resulting lag-frequency spectra and response function model fits are consistent with the results of section 3.




\end{document}